\documentclass{article}
\RequirePackage[a4paper,portrait,margin=30mm]{geometry}
\usepackage[svgnames,table]{xcolor}
\usepackage[utf8]{inputenc}
\usepackage[UKenglish]{isodate}
\usepackage{hyperref}
\usepackage{appendix}
\usepackage{graphicx}
\usepackage{todonotes}
\usepackage{tabularx}
\usepackage{booktabs}
\usepackage{caption}
\usepackage{ltablex}
\usepackage{setspace}%

\usepackage{fancyhdr}
\usepackage{datetime}

\title{An Evaluation of the Archive of Formal Proofs}
\author{
\small Carlin MacKenzie, Jacques Fleuriot and James Vaughan\\
\small Artificial Intelligence and its Applications Institute\\
\small School of Informatics\\
\small University of Edinburgh\\
{\small \{s1724780, jdf, s0952880\}@ed.ac.uk}}

\date{}

\begin{document}

\maketitle
\thispagestyle{empty}

\section{Introduction}
\label{sec:intro}

The Archive of Formal Proofs (AFP) is an online repository of formal proofs for the Isabelle proof assistant\footnote{\url{https://isabelle.in.tum.de}}. It first appeared on the internet in 2004, hosted as a static site on SourceForge\footnote{\url{https://afp.sourceforge.net}}, to serve as a central location for publishing, discovering, and viewing libraries of proofs. It subsequently took residence on its own domain\footnote{\url{https://isa-afp.org}}, however the visuals and functionality of the site have not been significantly updated since.

We conducted an online survey of the AFP in November 2020 to assess the suitability of the website. We first evaluated the questions for clarity and utility with a small pre-study group from the School of Informatics at the University of Edinburgh. The survey was then distributed to the Isabelle Mailing list\footnote{\url{https://lists.cam.ac.uk/mailman/listinfo/cl-isabelle-users}}, where it received 30 responses. \medskip

\noindent \textbf{Summary of results:} We found that long-term users of the website are generally satisfied with the Archive of Formal Proofs but that there are a number of areas, such as navigation, search and script browsing, that need improvement. 

\section{Pre-study}
\label{sec:intialdesign}

In order to validate the survey design, it was first distributed to the Artificial Intelligence Modelling Lab\footnote{\url{http://aiml.inf.ed.ac.uk}} in the School of Informatics at the University of Edinburgh. This group was chosen as the members are familiar with Isabelle across a variety of use cases and workflows.

\section{Survey Design}
\label{survey:design}

The pre-study was completed by 6 members of the Lab. After analysing their answers, we made the following adjustments to the survey, both to address limitations of the initial design and to account for the main audience i.e.\ the Isabelle mailing list.

\begin{itemize}
    \item The first question of the survey, ``Do you use the Archive of Formal Proofs?", was replaced with ``How often do you access the Archive of Formal Proofs?" as this was a more reliable way to filter out people who do not use the Archive.
    \item Two questions about submission were added. Although the submission form is separate from the AFP, it was felt that this would provide some further understanding of the responses.
    \item The final SUS question, ``I needed to learn a lot of things before I could get going with the AFP", was removed as the meaning was deemed to be ambiguous. To account for this in the SUS scoring, the rating of 3 was used for this question as it has no effect on the final score.
    \item Two questions about navigation regarding finding specific contents and entries by an author were added.
    \item Two questions about redesigning the UI and UX were added.
    \item Four questions changed their format from 5-star ratings to Likert scales \cite{likert32}, as these were found to be easier to analyse.
\end{itemize}

The survey was carried out via Microsoft Forms\footnote{\url{https://forms.office.com/}} and was organised as follows: 

\begin{enumerate}
  \item \textbf{Demographics:} 4 questions to filter users into different groups depending on their experience with the AFP.
  \item \textbf{Submission:} 2 questions to assess the submission process
  \item \textbf{System Usability Scale (SUS):} 9 SUS \cite{brooke1996sus} questions to act as an indicator of the usability of the AFP.
  \item \textbf{Biggest pain point:} 1 long answer question asking users to identify their biggest pain point.
  \item \textbf{Navigation:} 4 questions related to the ease of finding pages and page visit frequency.
  \item \textbf{Design:} 2 questions on the user interface and user experience.
  \item \textbf{Browsing session scripts:} 2 questions about the browsing experience and 1 short answer question about missing features.
  \item \textbf{Ranking priorities:} 1 question asking users to rank a number areas in order of importance.
\end{enumerate}

The final version of the survey can be found in Appendix~\ref{sec:surveyscript}.

\section{Participants}

The Isabelle mailing list was chosen as many of its members were likely to be users of the AFP, thereby increasing the likelihood of achieving a comprehensive evaluation of the website. The survey was advertised on the mailing list as \emph{Survey on the AFP} with an estimated time of 10-20 minutes. The time estimate was calculated based upon the average completion time of the pre-study. No compensation was advertised, and the main benefit of the study was to help guide some research on an evaluation of the AFP. The survey was open from 10 to 30 November 2020.

\newpage 

\section{Results and Analysis}
\label{sec:results}

In this section we look at the results more closely. We break these down to match the survey topics described in the previous section. 

\begin{figure}[ht]
    \centering
    \includegraphics[width=0.9\textwidth]{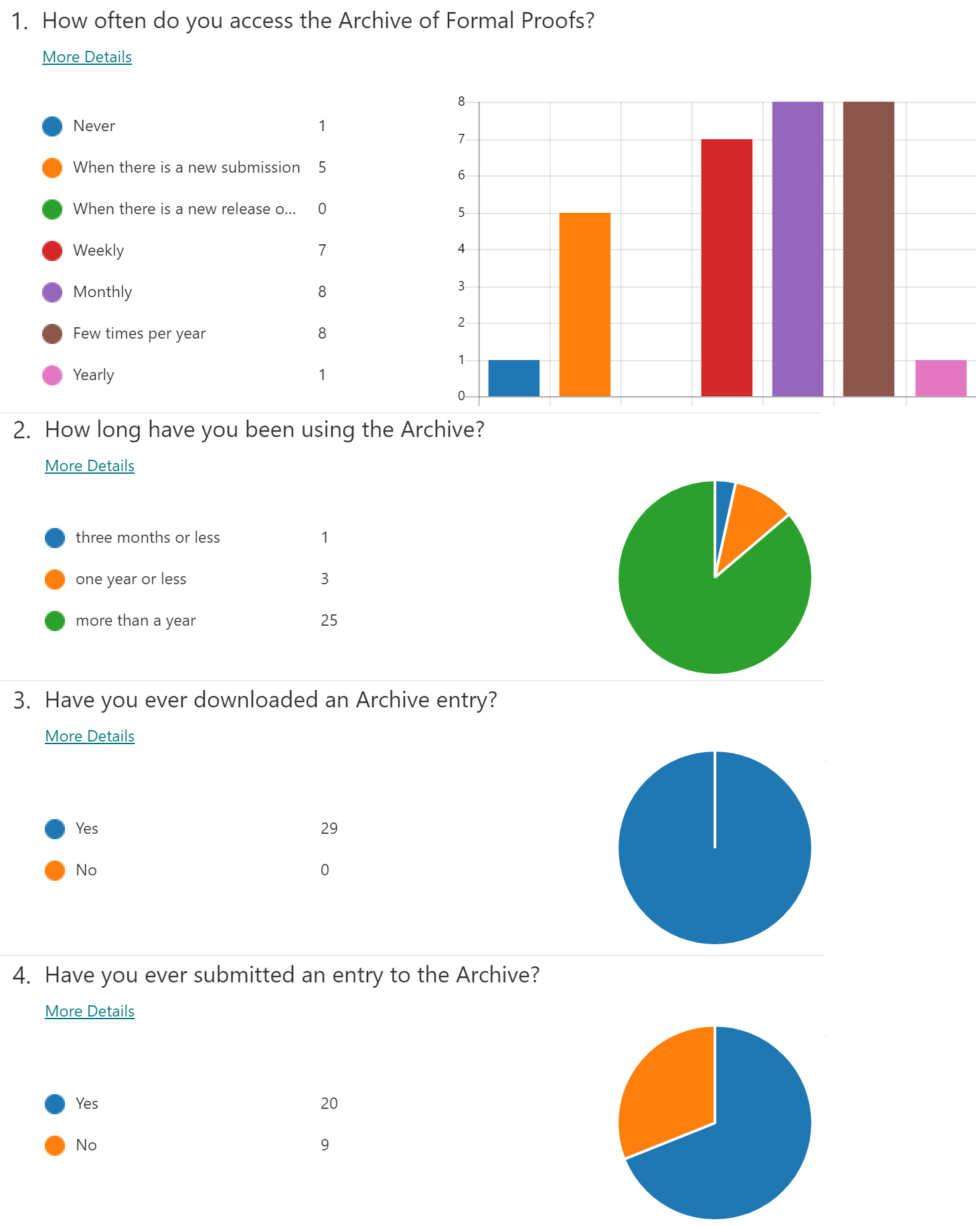}
    \caption{\textbf{Demographics.}
    The demographics of the respondents is skewed towards very active and long-term users.}
    \label{fig:demographics}
    \medskip
\end{figure}


\begin{figure}[ht]
    \centering
    \includegraphics[width=\textwidth]{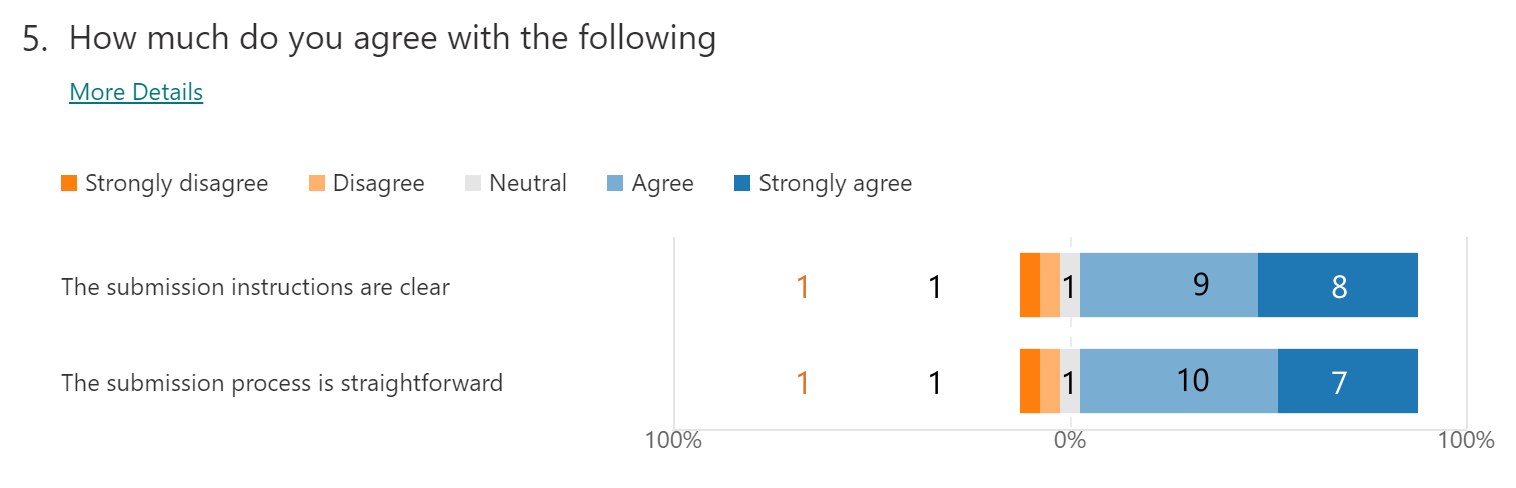}
    \caption{\textbf{Submission.} The vast majority of people who have submitted find the process clear and straightforward.}
    \label{fig:submission-1}
    \medskip
\end{figure}

\renewcommand{\arraystretch}{1.5}
\begin{table}[h!]
\centering
\rowcolors{5}{}{gray!10}
\begin{tabularx}{\textwidth}{X}
{\sf 6. What is your biggest pain point when submitting entries to the Archive?}
\vspace{0.3cm}\\ 
\hline
\footnotesize
Sometimes you get some errors from the system after submitting. And if I remember correctly, one time an entry didn't arrive because of a non-ASCII letter n an author name, but AFAIK this has been fixed now.\\
\footnotesize
Whether the entry will be accepted or not. \\
\footnotesize
In 2017, there was no ``preview" feature for the entry description.\\
\footnotesize
Converting apply-style proofs to Isar (not necessarily required by the AFP, but recommended)\\
\footnotesize
Forgetting to update something about a theorem before submission. \\
\footnotesize
Compared to a pull request on Github it is a bit more tedious and less transparent.\\
\footnotesize
Building of submission failing due to LaTeX issues without helpful error messages.\\
\footnotesize
Need to make sure the LaTeX part compile.\\
\footnotesize
To bring a submission into format. Sometimes this needs 5-6 times to make a submission attempt and to finally complete it.\\
\footnotesize
Having to run the new entry with the Isabelle development version if the new entry imports an entry which has been updated since the last release.\\
\footnotesize
Checking the Isabelle style rules. \\
\footnotesize
Getting the ROOT file right.\\ 
\hline
\end{tabularx}
\vspace{0.3cm} 
\caption{\textbf{Submission.}~Six of the comments were related to formatting of the ROOT file and script files. The most actionable feedback from this section was that error messages are often unhelpful and that there is no preview for the abstract section. Three participants had no discernible pain point with submission and are not included in the table}
\label{fig:submission-2}
\end{table}

\begin{figure}[h!]
    \centering
    \includegraphics[width=\textwidth]{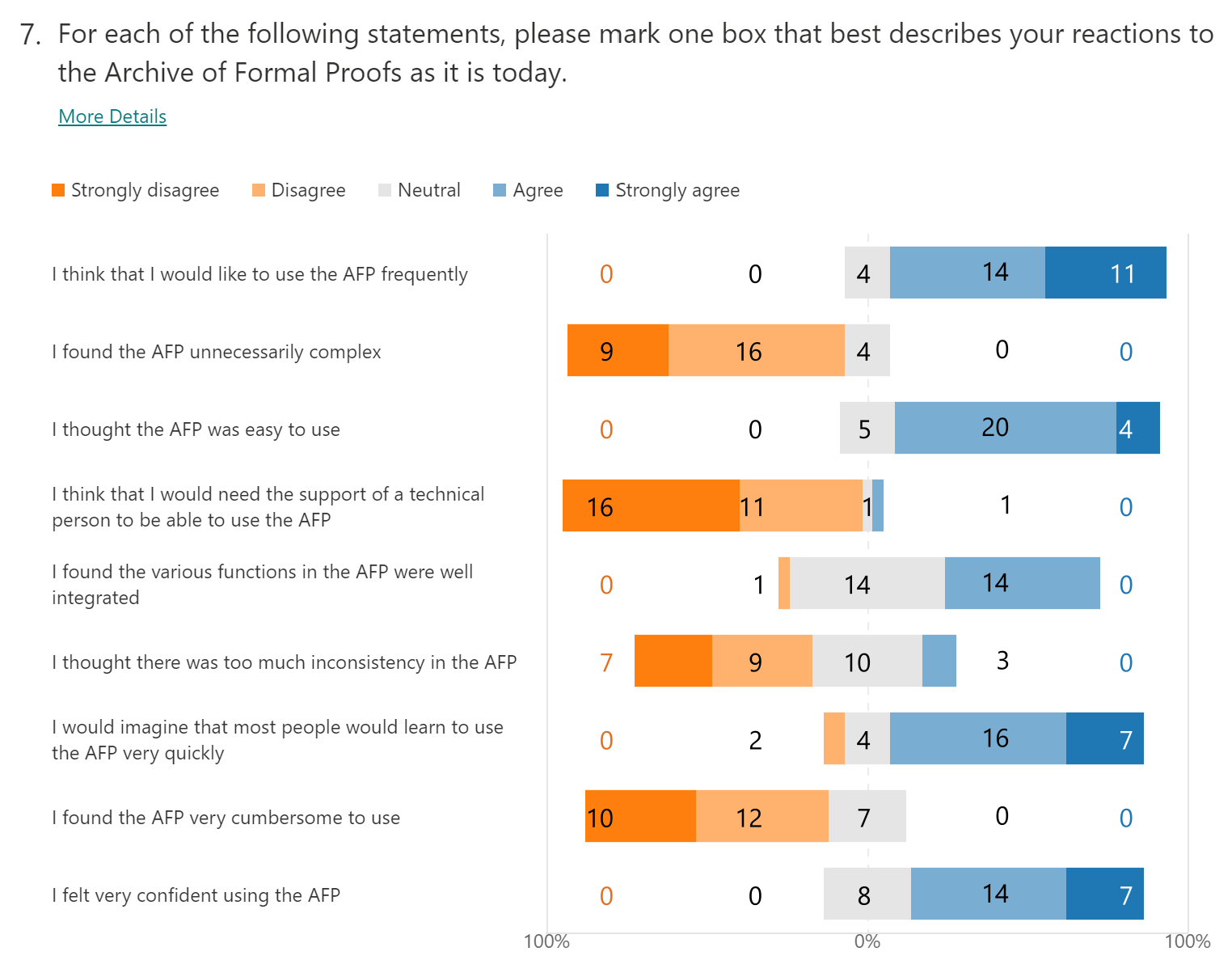}
    \caption{\textbf{SUS Questions.} The SUS score for the AFP is 72, which is above the average SUS score of 68 and suggests that the participants are satisfied by the AFP.}
    \label{fig:sus}
    \medskip
\end{figure}

\begin{table}[h!]
\centering
\rowcolors{5}{}{gray!10}
\begin{tabularx}{\textwidth}{X}
{\sf 8. What is your biggest pain point with the Archive? This could be with browsing entries, browsing scripts within entries, or any other feature of the AFP.}
\vspace{0.3cm}\\ 
\hline
\footnotesize
I think the biggest problem is that https://www.isa-afp.org/using.html is not explained well for Microsoft Windows.\\
\footnotesize
Use downloaded entries (e.g.integrate in a new development).  (This might be more an issue with Isabelle itself than AFP, I do not use Isabelle frequently.)\\
\footnotesize
Finding the correct Theory to import in Isabelle for a given Entry.\\
\footnotesize
I cannot online download and integrate the libs of AFP into Isabelle/HOL in the Isabelle/jedit UI.\\
\footnotesize
A lot of redundant formalizations (like graphs), making it non-obvious which to use.\\
\footnotesize
Problems with installing and using the new AFP version with every new Isabelle release.\\
\footnotesize
Rather weak HTML presentation.\\
\footnotesize
It's sometimes hard to find what you're looking for when you're just in search of ``a development that does X''.\\
\footnotesize
Learning what is there. As it grows, I do not know if my contributions are reinventing the wheel or if any theory for an entry in a different topic can help with my developments.\\
\footnotesize
The Proof Document contains all the proof, but the research value of the entry is usually in the published paper. A direct link would be useful.\\
\footnotesize
The scope could be clearer. In particular: What do I do with work in progress? Are many small libraries or one big library preferred? What about new tools, i.e. new tactics implemented in ML without any new proofs? How do I add a library from the AFP as a dependency to my project? (The method described at  https://www.isa-afp.org/using.html lacks basic functionalities like versioning or automatic downloading of dependencies and is a system wide setting instead of a per-project setting.)\\
\footnotesize
Searching if a theory already does something I need.\\
\hline
\end{tabularx}
\vspace{0.3cm}
\caption{\textbf{Biggest Pain Point.}~The most common response was problems using AFP entries with Isabelle/jEdit \cite{wenzel2012isabelle}. In total, 6 people had this problem, from lacking instructions for Windows to finding the correct theory to import from an entry. The next largest area was search, with 3 respondents describing issues relating to finding whether there is an entry which does what they need. There were four specific asks: one would like a direct link to the corresponding paper about the entry if applicable; another finds the HTML presentation weak; yet another finds it difficult to choose between many similar entries; finally, one user is confused of the scope of the project and what entries are worthy of entering. Finally, 3 respondents had no pain point with the AFP and their responses are not included in the table.}
    \label{fig:pain-point}
\end{table}

\clearpage 

\begin{figure}[h!]
    \centering
    \includegraphics[width=\textwidth]{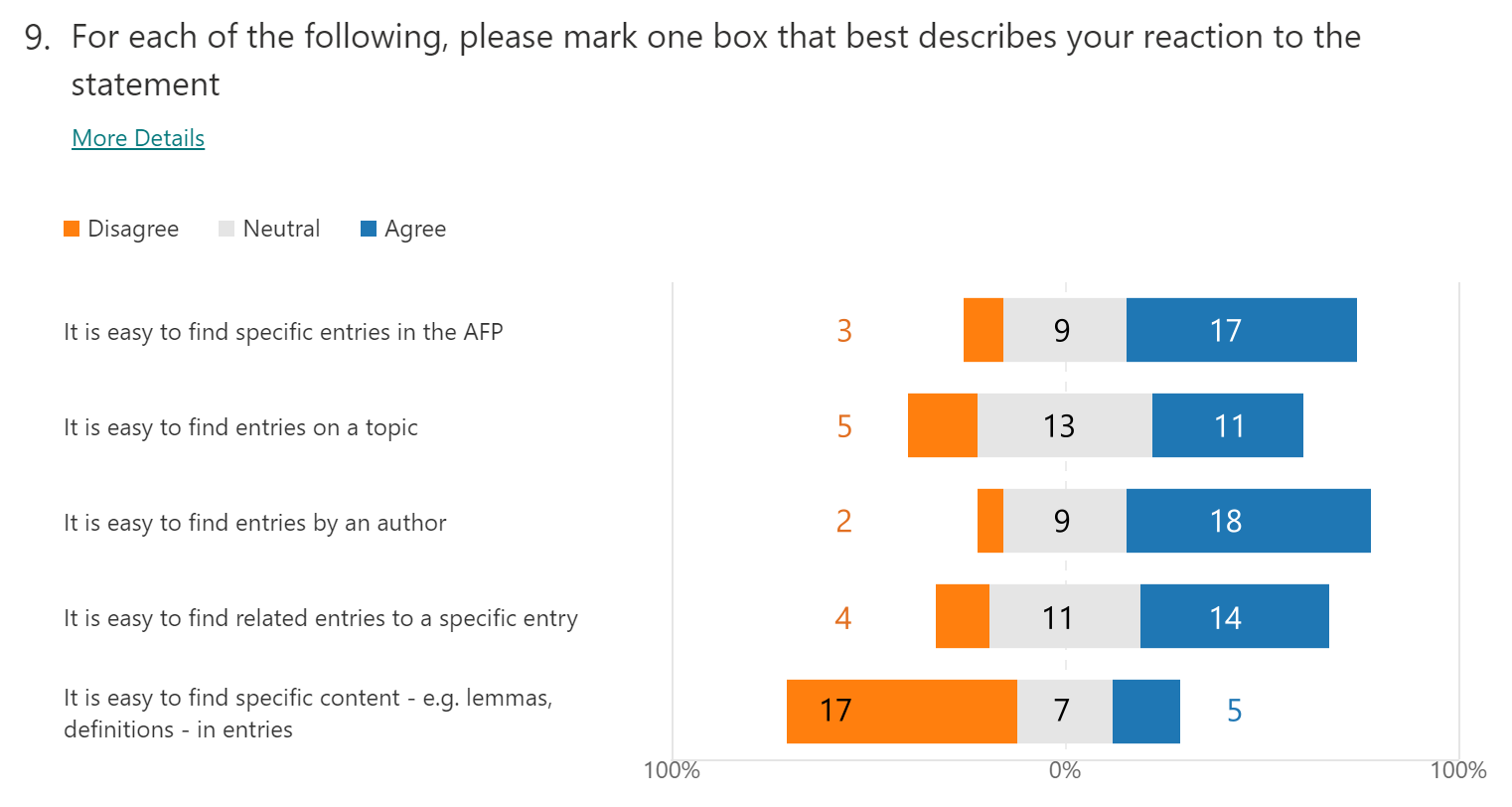}
    \caption{\textbf{Navigating to Specific Content.}
    Most content is easy to navigate to, except for specific content in entries. Notably, there is no category in which everyone is neutral or agrees implying that navigation can be improved in all areas.}
    \label{fig:navigating-to-specific-content}
    \medskip
\end{figure}


\begin{figure}[t]
    \centering
    \includegraphics[width=\textwidth]{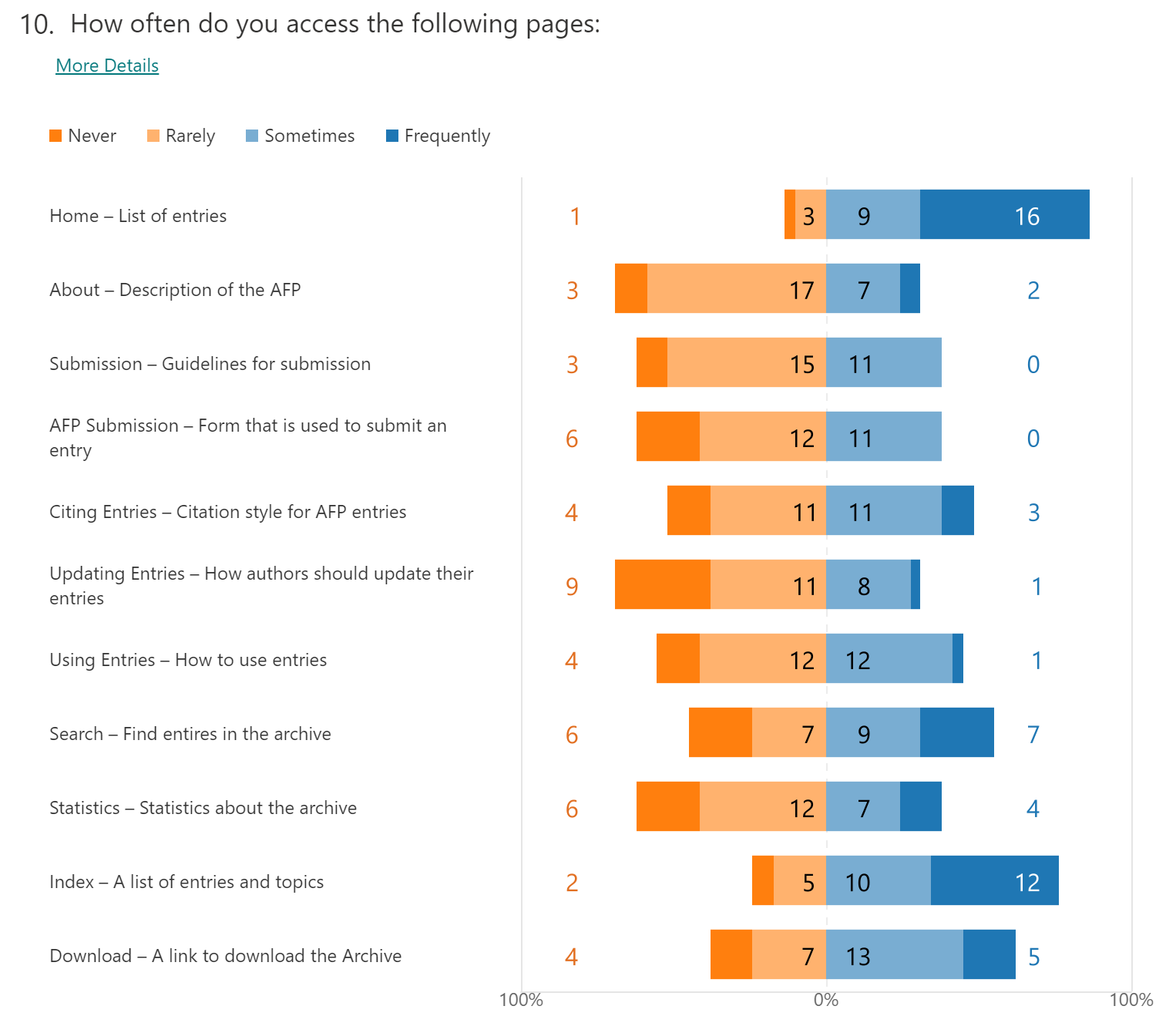}
    \caption{\textbf{Navigating to Pages.}
    There are very different access requirements for pages of the AFP even though all but two of the eleven pages feature in the sidebar.}
    \label{fig:navigating-to-pages}
    \medskip
\end{figure}


\begin{figure}[h]
    \centering
    \includegraphics[width=\textwidth]{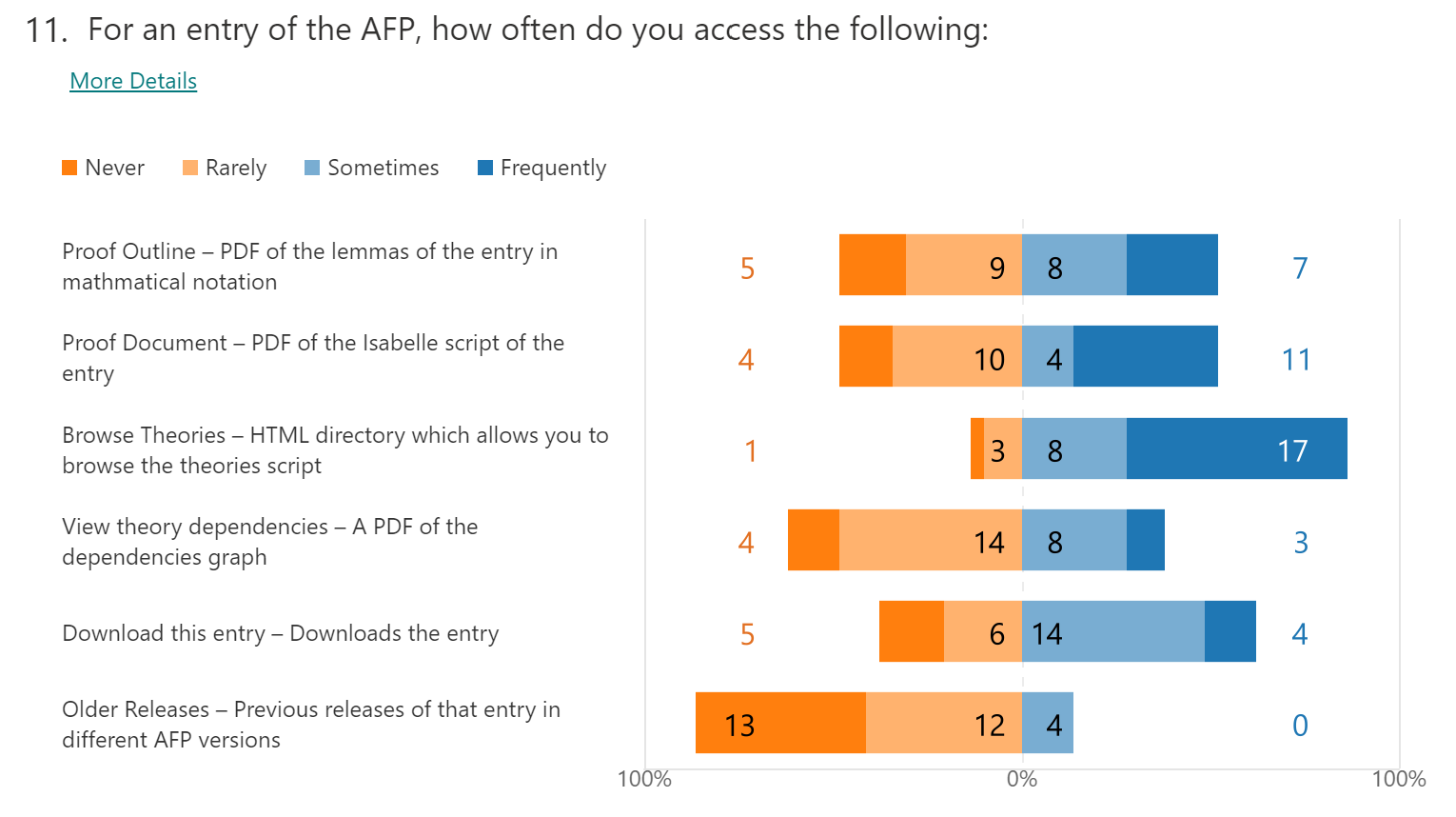}
    \caption{\textbf{Navigating to Pages Related to the Entry.}
    Each entry of the AFP has several links to pages related to it. ``Browse Theories" and ``Download" are the most accessed while ``Older Releases" is rarely accessed.}
    \medskip
\end{figure}


\begin{figure}[h]
    \centering
    \includegraphics[width=\textwidth]{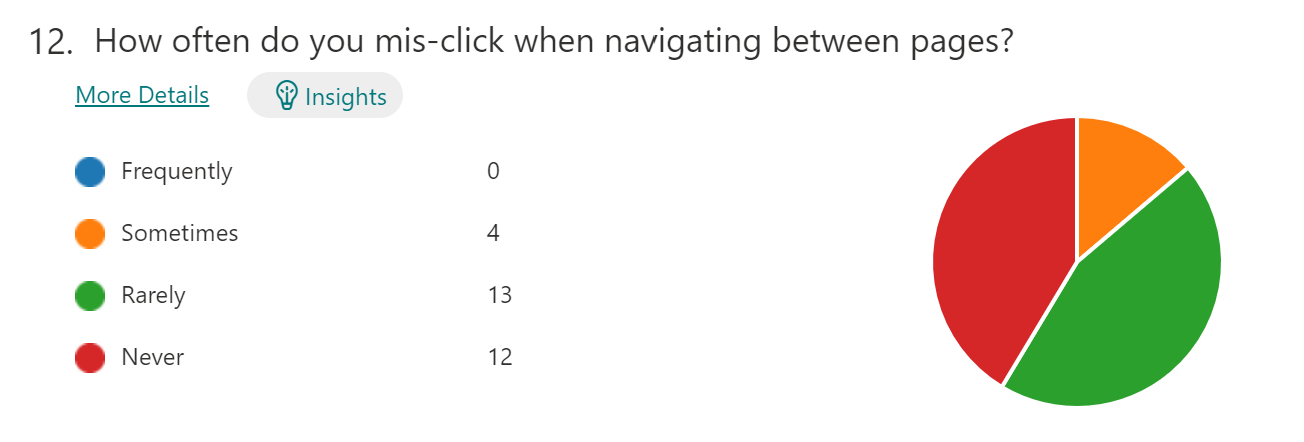}
    \caption{\textbf{Clarity of Link Text.}
    Over half the participants mis-click rarely or sometimes.}
    \label{fig:clarity-of-link-text}
    \medskip
\end{figure}

\begin{figure}[h]
    \centering
    \includegraphics[width=\textwidth]{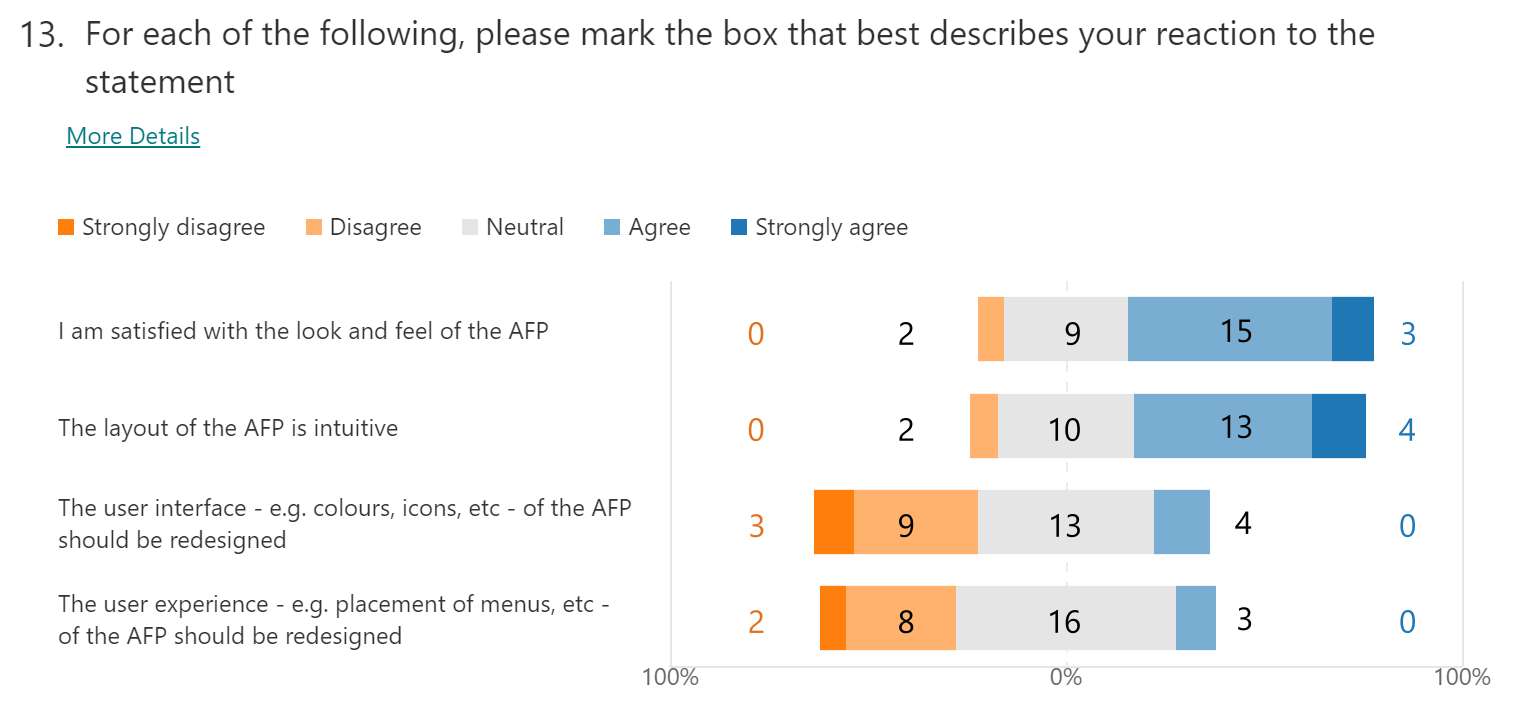}
    \caption{\textbf{Design and User Experience.}
    Most users are satisfied with the UI and UX but are neutral towards a redesign of either.}
    \medskip
\end{figure}


\begin{figure}[h]
    \centering
    \includegraphics[width=\textwidth]{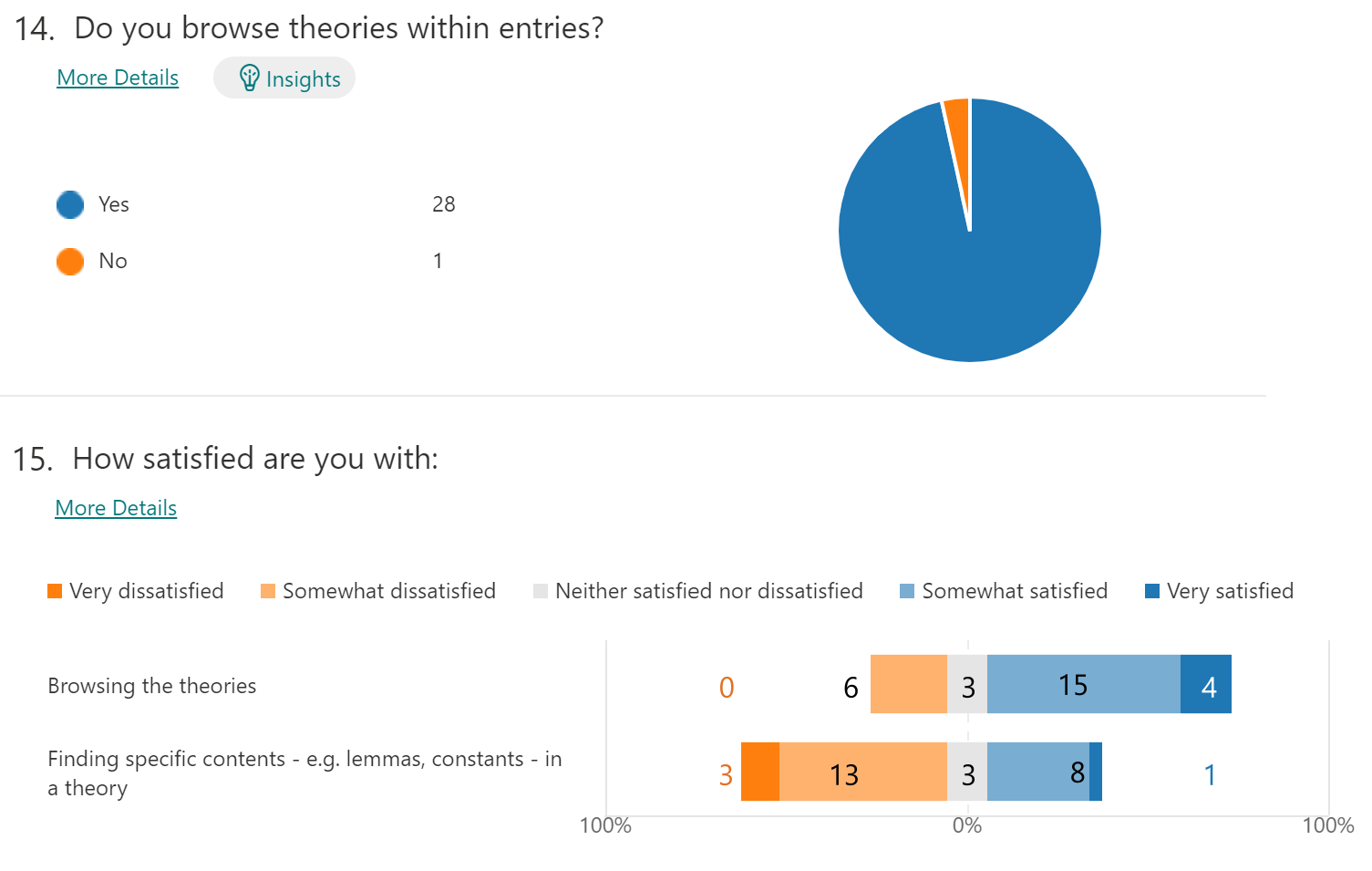}
    \caption{\textbf{Browsing Theories.}
    Almost all participants browse theories and are mostly satisfied with the experience. However, they are largely unsatisfied with finding contents within theories.}
    \label{fig:theory-scripts-1}
    \medskip
\end{figure}

\begin{table}[h!]
\centering
\rowcolors{5}{}{gray!10}
\begin{tabularx}{\textwidth}{X}
{\sf 16. What feature would improve browsing theory scripts?}
\vspace{0.3cm}\\ 
\hline
\footnotesize
The ctrl-click/cmd-click option of JEdit to find theorems and constants available in the online version.\\
\footnotesize
If I could navigate to the definition of a type or a constant by clicking on it.\\
\footnotesize
Summary/Outline Feature. Goto Definition/Usage Statistics about frequently used theorems.\\
\footnotesize
Search for a lemma and a definition. Click \& jump like in jEdit when navigating theory.\\
\footnotesize
Linking https://search.isabelle.in.tum.de/ would improve the search experience.\\
\footnotesize
More structure and links in the HTML output.\\
\footnotesize
Links from entities to where they are defined or proved.\\
\footnotesize
Index of lemmas.\\
\footnotesize
A proper search function. \\ 
\footnotesize
A Sidekick of the theory.\\
\footnotesize
Semantic search.\\
\footnotesize
Ontology and ontology based search.\\
\footnotesize
I don't know\\
\footnotesize
Maybe something like ``sidekick" from Isabelle/jEdit. Maybe a better search.\\
\footnotesize
Clickable terms with a link that leads to the definition!!! that would be awesome!; Crossreferences; overlays that show information about terms.\\
\footnotesize
Add some features from jEdit: highlighting of inner syntax, go to definition hyperlinks, search theorems and search constants functionality, text search across all files. Also: option to find all uses of a   constant or lemma.\\
\hline
\end{tabularx}
\vspace{0.3cm}
\caption{\textbf{Browsing Theory Scripts.}
    16 people responded to this question and half of them requested the ability to be able to click on links to definitions, as available in Isabelle/jEdit. Following this was 7 requests for better search capabilities and 5 requests for SideKick functionality (an outline of the sections, lemmas, etc). One respondent suggested usage statistics of frequently used theorems.}
    \label{fig:theory-scripts-2}
\end{table}

\begin{figure}[h]
    \centering
    \includegraphics[width=\textwidth]{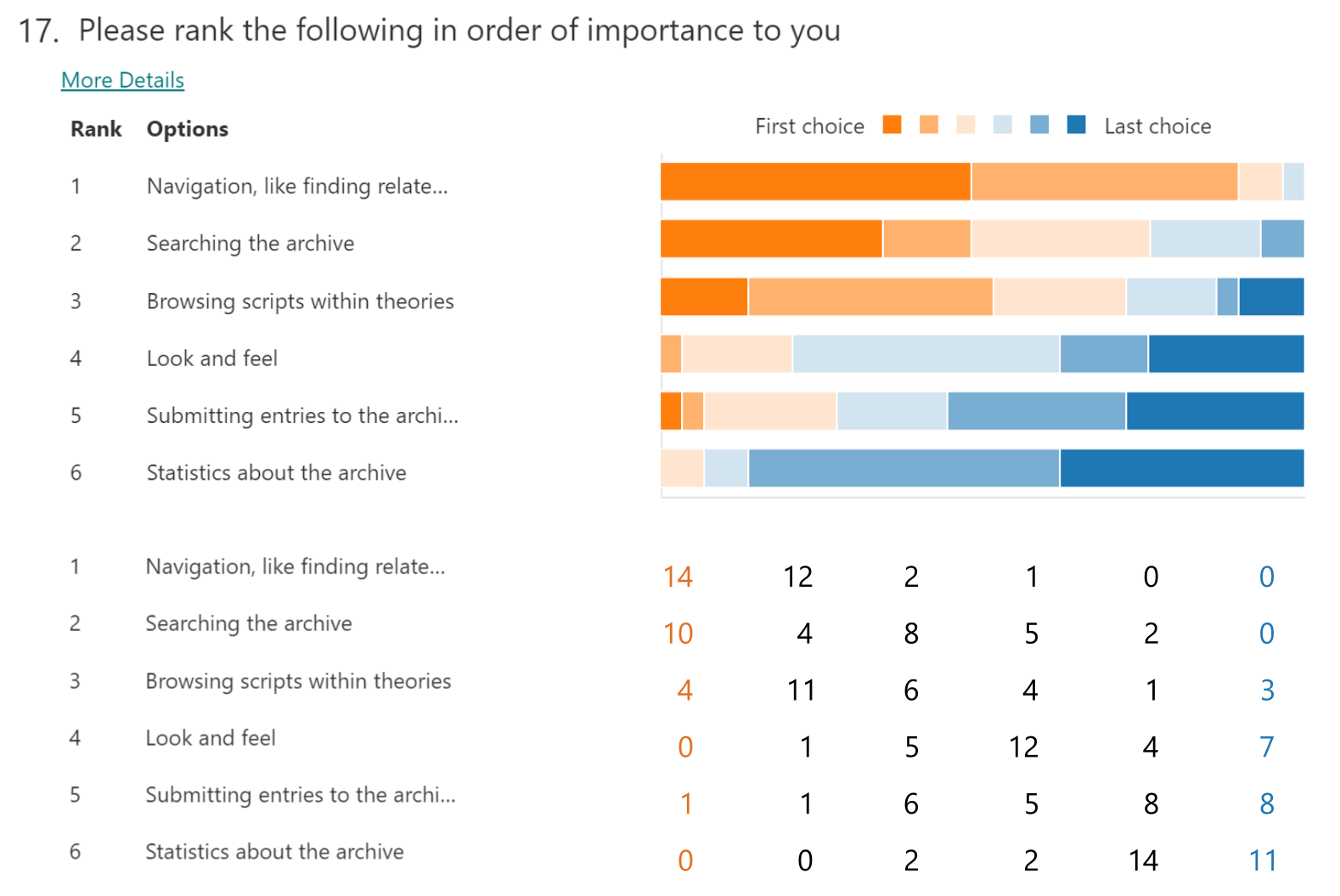}
    \caption{\textbf{Ranking Priorities.}
    The ordering of priorities is consistent across the participants. Look and feel is a low priority which is congruous with the neutrality towards a redesign.}
    \label{fig:ranking-priorities}
    \medskip
\end{figure}

\clearpage

\section{Summary}
\label{sec:analysis}


It is likely that people who are subscribed to the Isabelle mailing list and willing to answer the survey  are active AFP users. This is reflected in the demographics and the familiarity of the audience should be kept in mind when interpreting the results.\smallskip

The survey was taken by 30 participants and 29 of them answered all the parts. From Fleuriot et al.\ \cite{fleuriot2016social}, there were around 600 contributing users on the mailing list in the seven year period of 2008 to 2015. We cannot tell whether the mailing list has grown or shrunk in the years since, but the number of responses seems adequate for the order of magnitude of the mailing list.\smallskip

As the participants are mostly contributors to the AFP, their opinions are highly valued. The SUS score of 72 implies that they are generally satisfied with the AFP, which is a testament to the design decisions that have lasted almost 20 years. The pre-study score was much lower, 46, which seems to imply that non-contributors of entries to the AFP might be less satisfied. However a further study with a larger group would be needed to confirm this.\smallskip

Three respondents described difficulty in finding existing functionalities and seven requested improvements to script search capabilities. Additionally, a majority of the participants struggle to find specific content in the AFP. As it is the second highest priority for users, it is clear that the AFP would be more useful if the search capabilities were improved.\smallskip

The most important thing for participants was navigation and the results of the survey imply that it does not currently meet their needs. The sidebar is the main navigation area and it is not ordered in terms of frequency of use (Figure~\ref{fig:navigating-to-pages}), audience (contributor \emph{vs} non-contributor) or content (i.e.\ ``Home" and ``Index" are the only pages which list entries and they are separate). Participants also report mis-clicking, which could suggest link labelling is not clear or links are too small. It is also hard to find many different types of content as shown in Figure~\ref{fig:navigating-to-specific-content}. Navigation is closely related to search, however, and many of these issues could be solved in other ways. \smallskip

Finally, navigation improvements to script browsing were requested frequently -- over half the respondents requested in-place links to entity definitions, i.e.\ to directly navigate to specific content. Similarly having an outline of the theory file, as SideKick provides in Isabelle/jEdit, was highly requested.

\section{Conclusion and Further Work}
\label{sec:conclusion}

Whilst a significant minority of responses hold that redesign is unnecessary, there are many specific criticisms with the current design as well as a general sentiment that several core features (specifically navigation, search, and theory browsing) could be improved. \smallskip

Future work could broaden the evaluation to users who are unfamiliar with the AFP, to understand how first-time users understand the functionality of the website. This would be important to understand as it is crucial to the longevity of the project that new people can be on-boarded successfully. 

\section*{Acknowledgement}

We would like to thank all those who took part in this survey of the Archive of Formal Proofs. Our thanks also go to the developers and maintainers of the AFP for providing such a valuable resource to the theorem proving community. 

\bibliography{main}
\bibliographystyle{ieeetr}

\appendix  

\section{Survey Script}
\label{sec:surveyscript}

\begin{figure}[h!]
    \centering
    \includegraphics[width=0.9\textwidth]{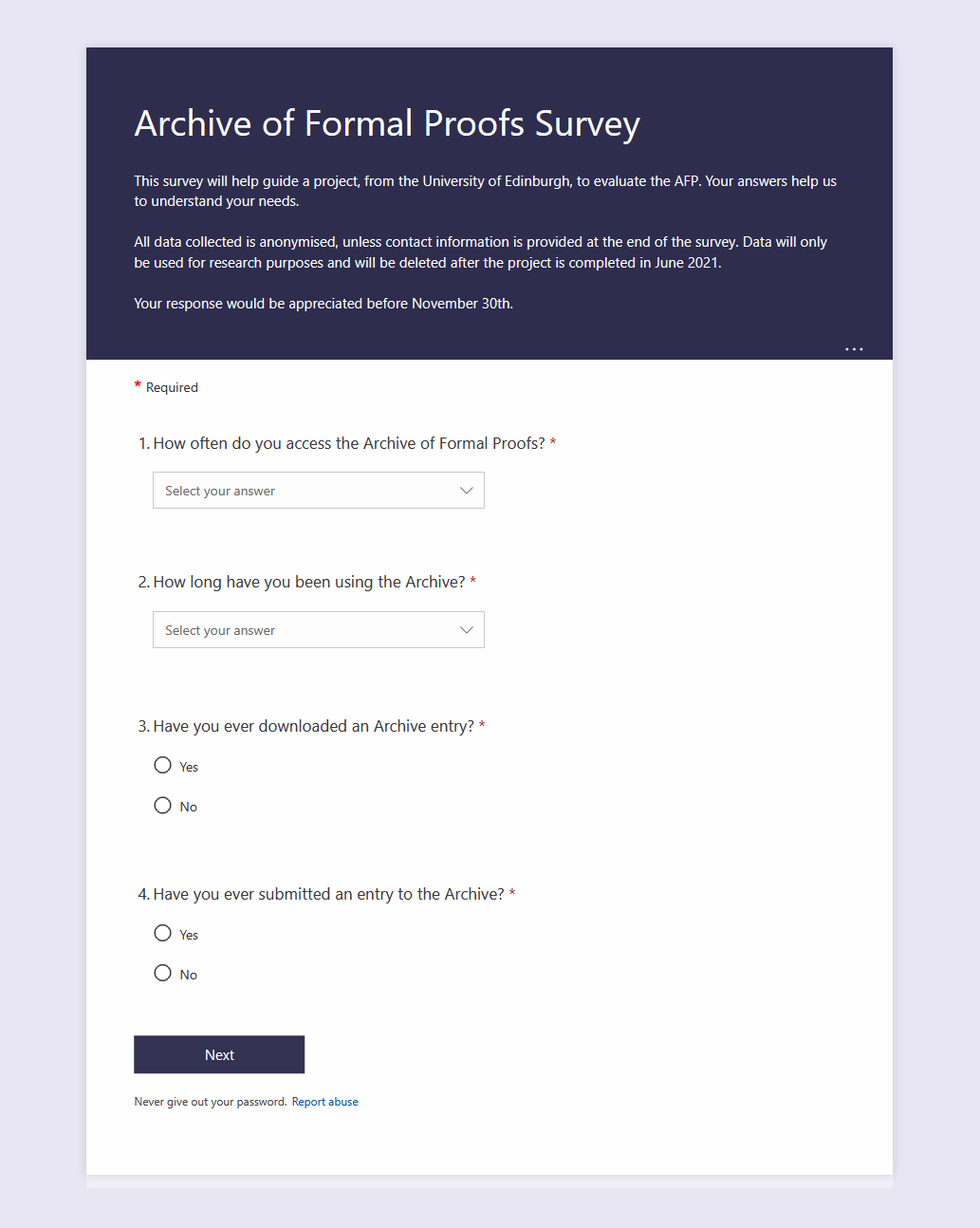}
\end{figure}

\begin{figure}[h]
    \centering
    \includegraphics[width=\textwidth]{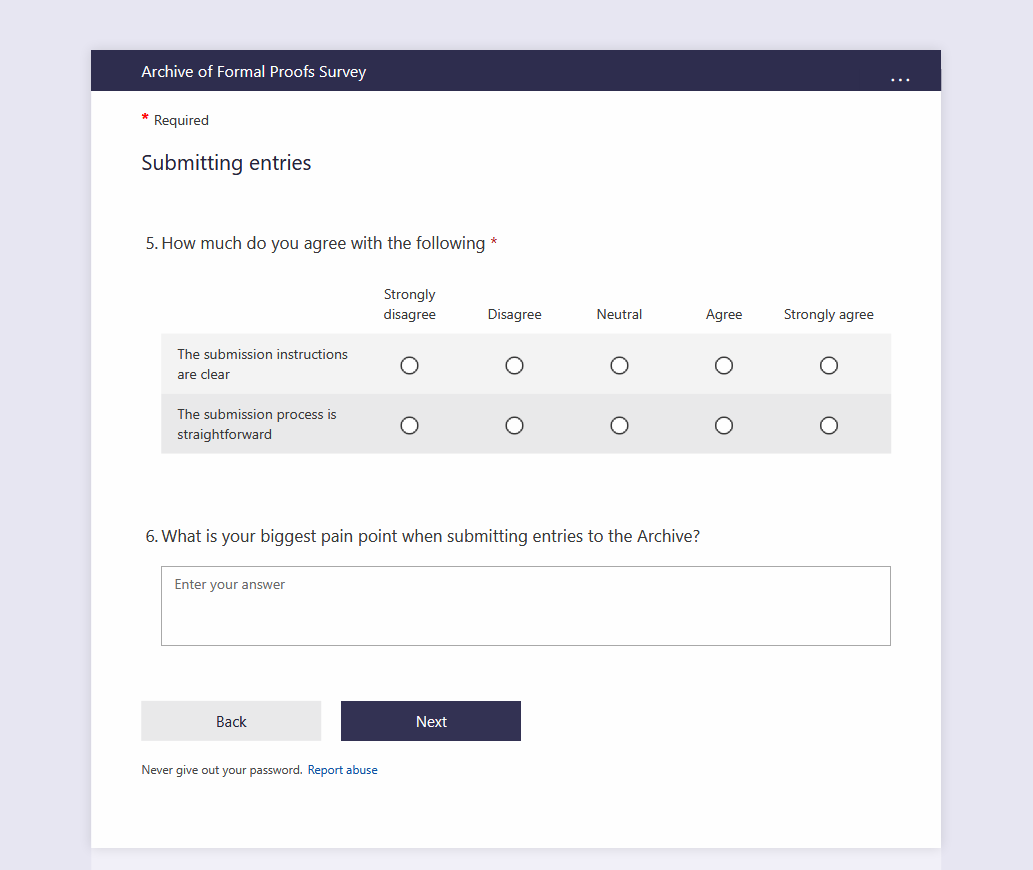}
\end{figure}

\begin{figure}[h]
    \centering
    \includegraphics[width=\textwidth]{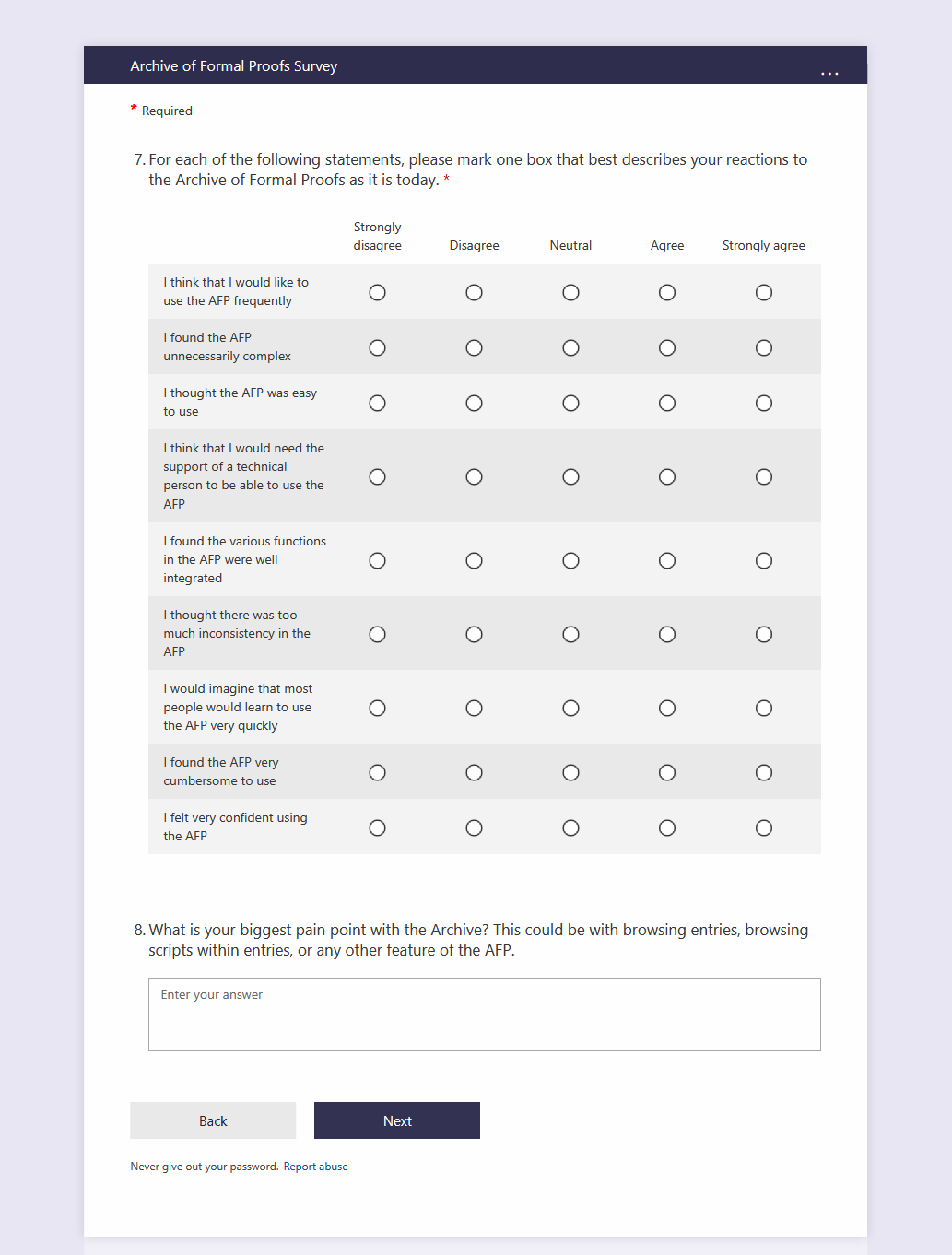}
\end{figure}

\begin{figure}[h]
    \centering
    \includegraphics[width=\textwidth]{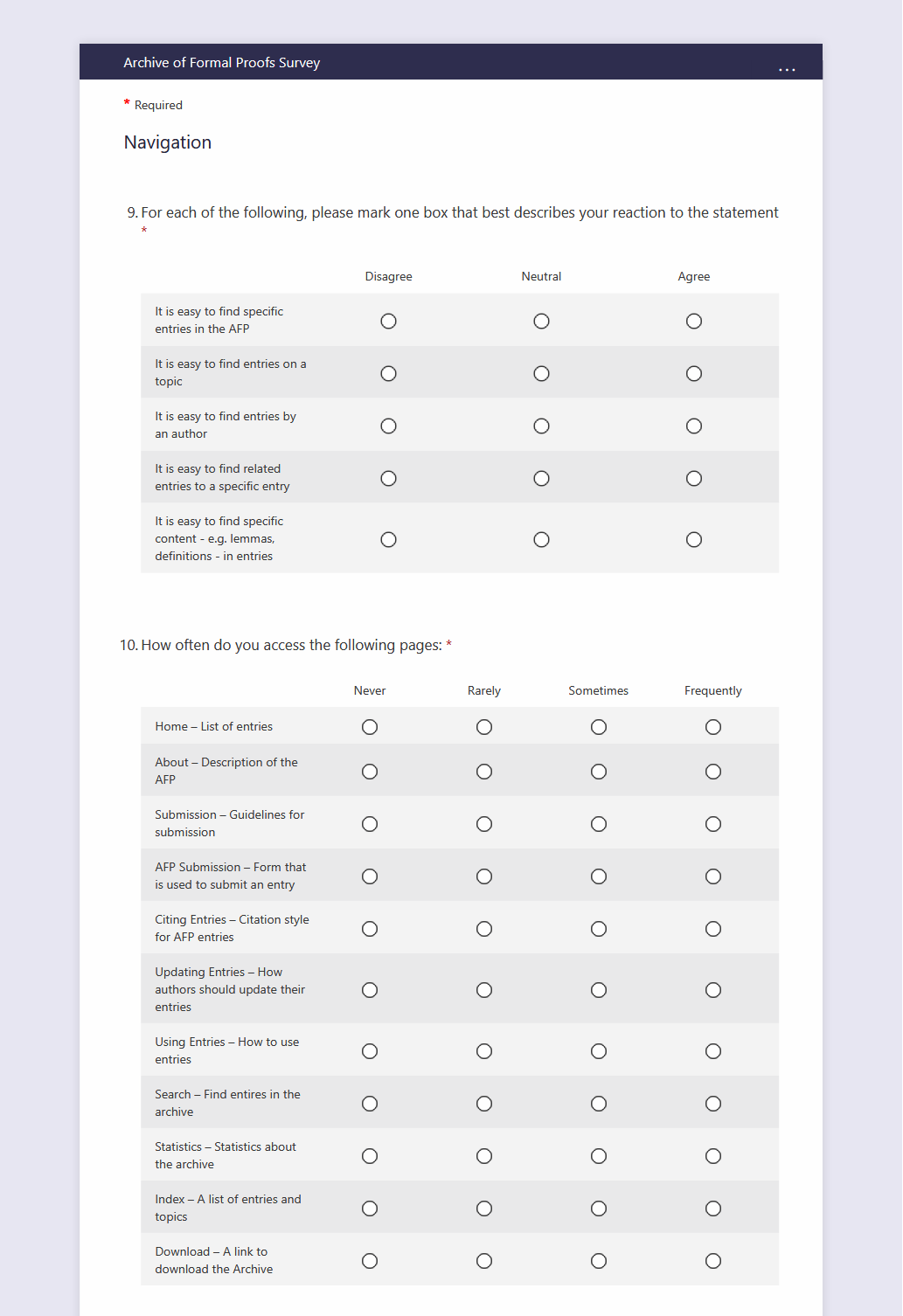}
\end{figure}

\begin{figure}[h]
    \centering
    \includegraphics[width=\textwidth]{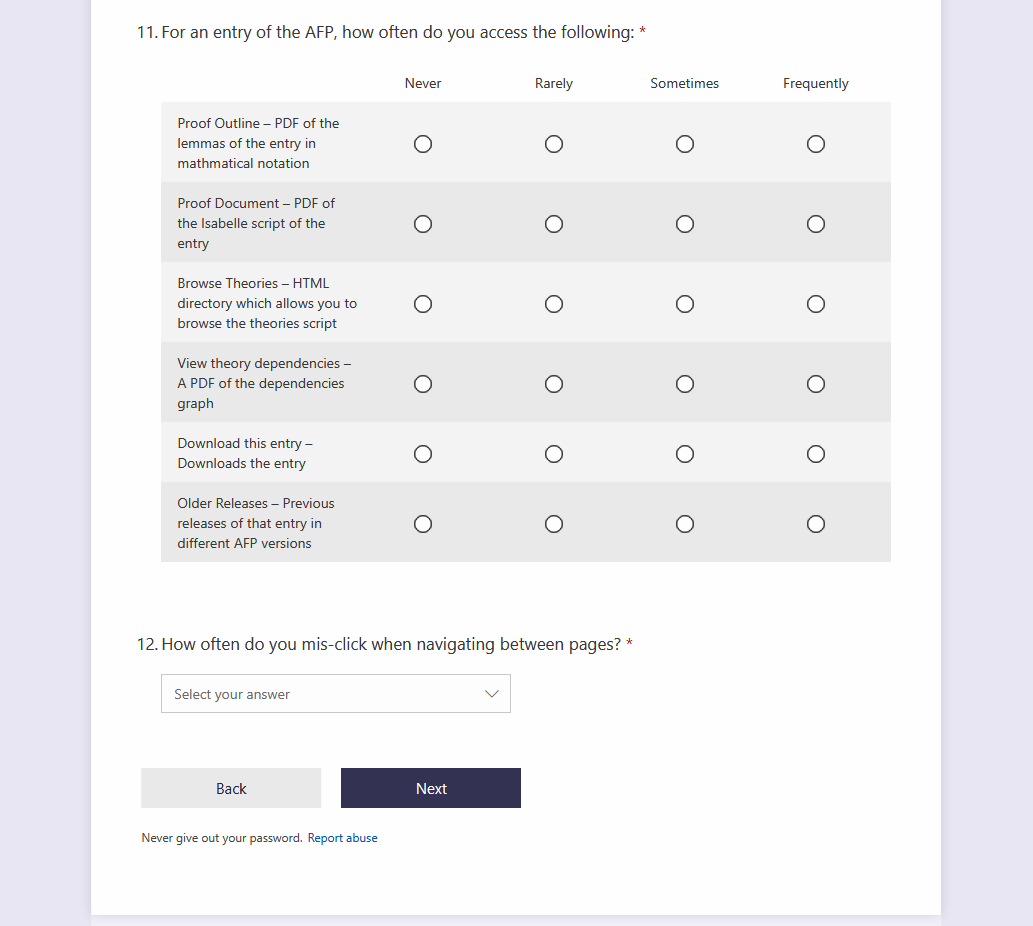}
\end{figure}

\begin{figure}[h]
    \centering
    \includegraphics[width=\textwidth]{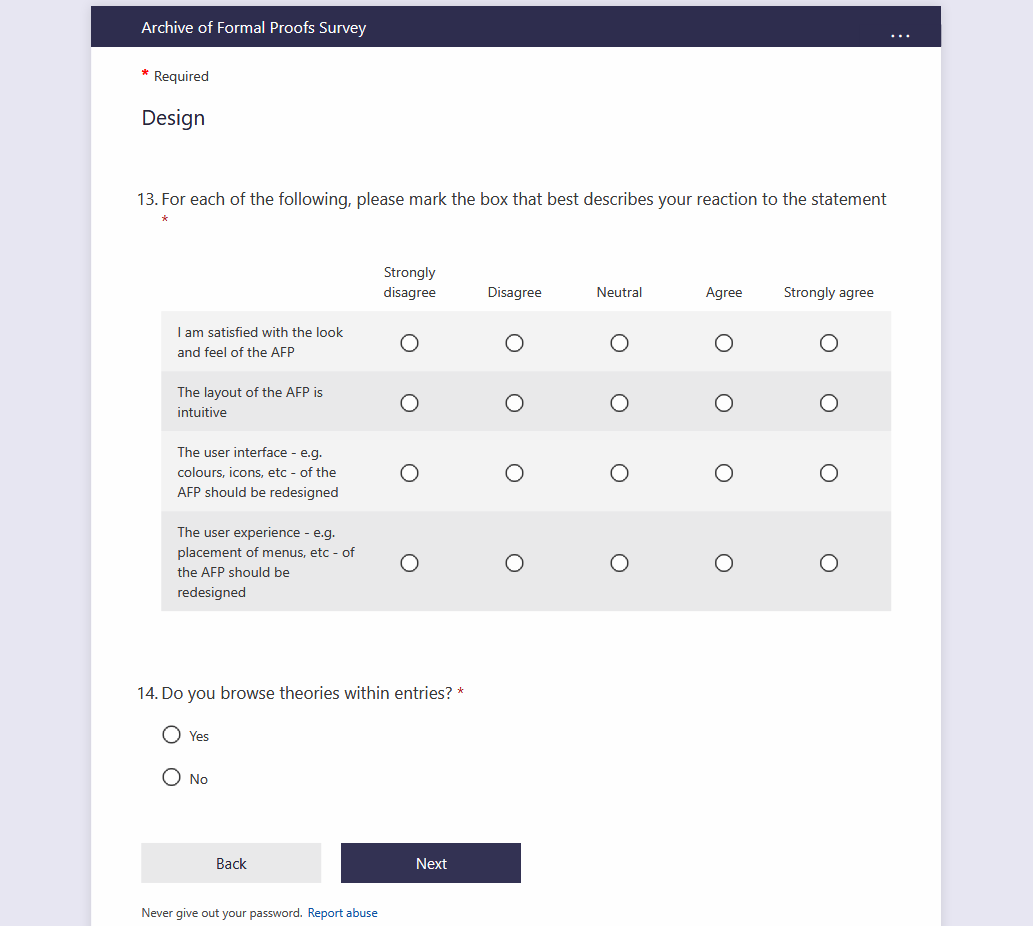}
\end{figure}

\begin{figure}[h]
    \centering
    \includegraphics[width=\textwidth]{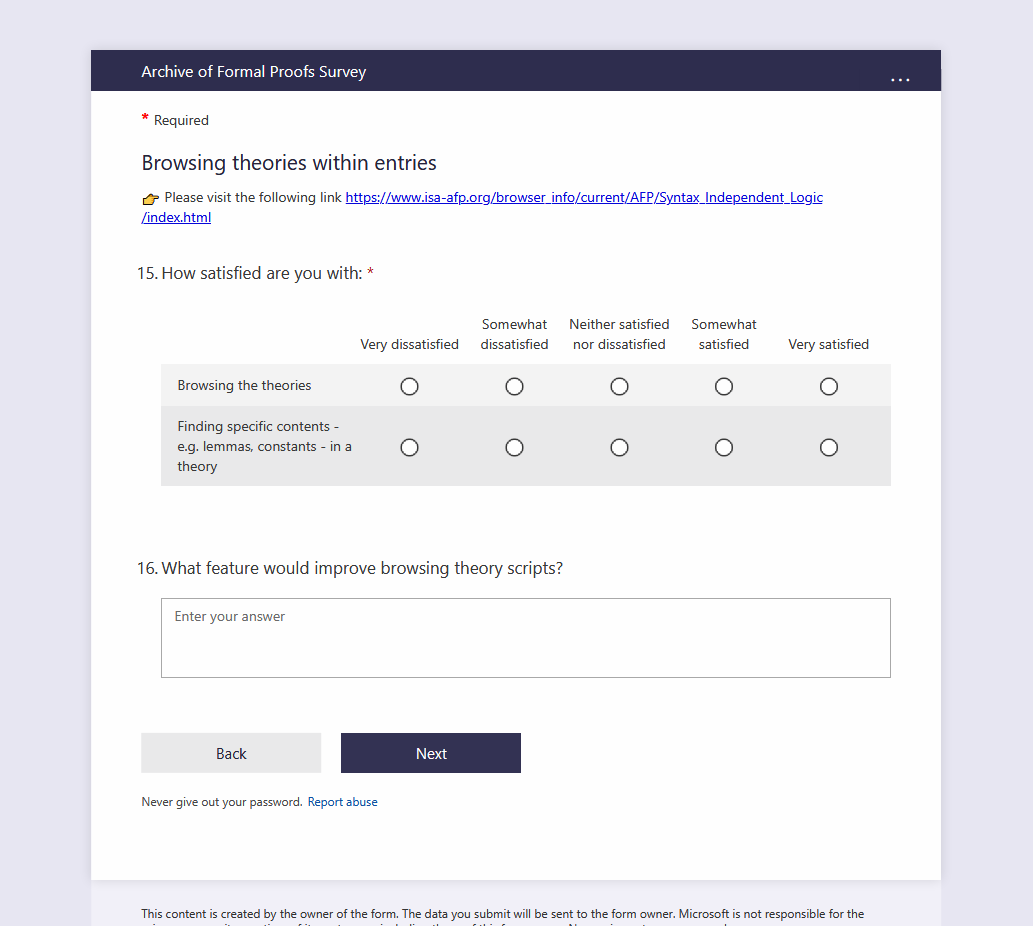}
\end{figure}

\begin{figure}[h]
    \centering
    \includegraphics[width=\textwidth]{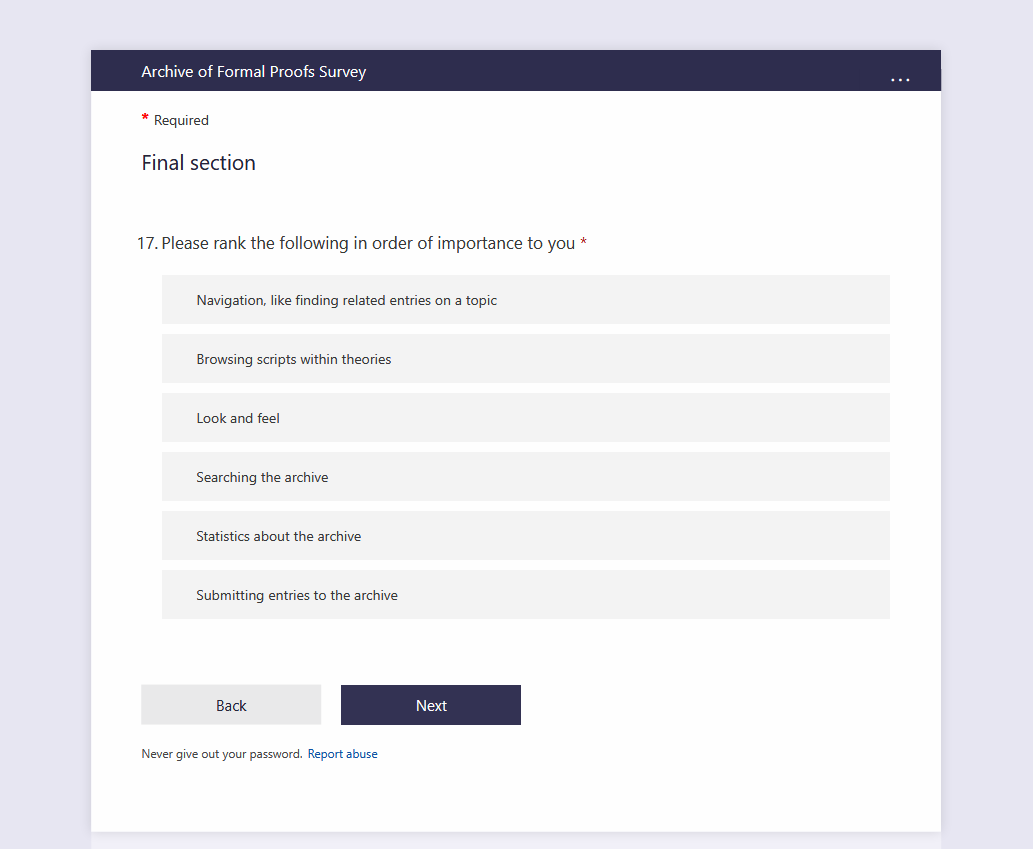}
\end{figure}

\end{document}